\documentclass[final,1p,times,twocolumn]{elsarticle}
 
\usepackage{amssymb}

\usepackage{amsthm}
\usepackage{amsmath}
\usepackage{booktabs}

\journal{Chaos, Solitons \& Fractals}

\begin{document}

\begin{frontmatter}

\title{Modeling the effect of the vaccination campaign on the Covid-19 pandemic}

\author[1]{Mattia Angeli\corref{cor1}}
\ead{mangeli@seas.harvard.edu}
\author[1]{Georgios Neofotistos}
\author[1]{Marios Mattheakis}
\author[1,2]{Efthimios Kaxiras}

\cortext[cor1]{Corresponding author}
\address[1]{John A. Paulson School of Engineering and Applied Sciences,
Harvard University, Cambridge, Massachusetts 02138, USA}
\address[2]{Department of Physics, Harvard University, Cambridge, Massachusetts 02138, USA}

\begin{abstract}
Population-wide vaccination is critical for containing the SARS-CoV-2 (Covid-19) pandemic
when combined with restrictive and prevention measures.
In this study we introduce SAIVR, a mathematical model able to forecast the Covid-19 epidemic evolution during the vaccination campaign.
SAIVR extends the widely used Susceptible-Infectious-Removed (SIR) model by considering the Asymptomatic (A) and Vaccinated (V) compartments.
The model contains several parameters and initial conditions that are estimated by employing a semi-supervised machine learning procedure. 
After training an unsupervised neural network to solve the SAIVR  differential equations, a supervised framework then estimates the optimal conditions and parameters that best fit recent infectious curves of 27 countries.
Instructed by these results, we performed an extensive study on the temporal evolution of the pandemic under varying values of roll-out daily rates, vaccine efficacy, and a broad range of societal vaccine hesitancy/denial levels.
The concept of herd immunity is questioned by studying future scenarios which involve different vaccination efforts and more infectious Covid-19 variants.

\end{abstract}


\end{frontmatter}


\section{Introduction}
\label{sec:intro}
The World Health Organization (WHO) declared the SARS-CoV-2 (Covid-19) outbreak in Wuhan to be a pandemic on March 11, 2020.
Since then, Covid-19 has become a serious global health threat due to its rapid spread, transmission through asymptomatic infected individuals and complex epidemiological dynamics.
As of May 2021, already more than 3 million lives have been lost due to the virus. The spread of SARS-CoV-2 has thus far been extremely difficult to contain. \\
By the end of 2020, the successful development of effective vaccines and the onset of their widespread distribution in most of the world's countries, was hailed as the decisive mean to contain the pandemic.
However, important questions linger on whether the vaccination effort will succeed in effectively eradicating the disease.
The appearance and wide spread of more contagious SARS-Cov-2 strains, the onset and scale of the vaccine deployment and high levels of vaccine hesitancy/denial in the society, are among the key factors hindering the vaccination effort and the achievement of herd immunity. 
Modeling the impact of these key factors on the  evolution of the pandemic is of critical importance for assessing the vaccination effectiveness against it.\\
In studying past epidemics, scientists have systematically applied ''random mixing'' compartmental
models which assume that an infectious individual can spread the disease to any susceptible member of the population before becoming recovered or removed, 
as originally considered by Kermack and McKendrick~\cite{Kermack_1927}. These models constrain the total population in compartments by considering stages of the infection and flows among them.

In the present study we propose a new model named SAIVR, which incorporates two important characteristics of the Covid-19 epidemic, namely the considerable transmission of the disease by asymptomatic infected individuals and the vaccination campaign with World Health Organization (WHO) approved vaccines. 
More recent modeling approaches involve agent-based simulations ~\cite{Kaxiras_2020}, heterogeneous social networks ~\cite{Barthelemy_2005, Ferrari_2006, Volz_2008, Tagliazucchi_2020, Vespignani_2018, Zhang_2016}, and Bayesian inference models \cite{Groendyke_2011}.
Although a large number of research studies are currently investigating the Covid-19 epidemiological characteristics ~\cite{Sanche_2020, Li_2020,Imai_2020, Rothe_2020,  Wynants_2020, Koh2020, Riccardo2020, khalili2020, Jie2020}, we believe that a simple but efficient model, which can capture the basics of the complex behavior of the pandemic including the vaccine roll-out, can offer useful guidance for the pandemic's near-term and longer-term evolution. 
By using a recently developed semi-supervised machine learning approach \cite{Flamant, Marios_2020, Paticchio_NIPS} we  systematically reproduced the pandemic dynamics during the 2021 spring in several different countries.  
We then used the model to assess the importance of a rapid vaccination campaign to prevent future outbreaks driven by more infectious variants.
\\

The work is organized as follows. In Sec.~\ref{sec:SAIVR} we introduce the SAIVR model and its parameters. The machine learning approach that we used to reproduce the infectious curves of 27 selected countries/states is thoroughly described in Sec.~\ref{sec:NN}. In Sec.~\ref{sec:future_scenarios} we study future scenarios involving more infectious variants making quantitative arguments on how they might affect herd immunity.
Sec.~\ref{sec:Conclusions} is devoted to concluding remarks.

\section{The SAIVR model}  \label{sec:SAIVR}

One of the first attempts to mathematically describe the spread of an infectious disease is due to Kermack and McKendrick~\cite{Kermack_1927}. In 1927 they introduced the so-called Susceptible-Infectious-Removed (SIR) model.
The SIR model describes the dynamics of a (fixed) population of $N$ individuals split into three compartments:
\begin{itemize}
\item
$S(t)$ is the Susceptible compartment that counts the number of individuals susceptible but still not infected by the disease;
\item
$I(t)$ is the Infectious compartment that counts the number of infectious individuals;
\item
$R(t)$ is the Removed compartment. It represents the number of those who can no longer be infected either because they recovered and gained long-term immunity or because they passed away.
\end{itemize}

\begin{figure*}[h]
\begin{center}
\includegraphics[width=0.95\textwidth]{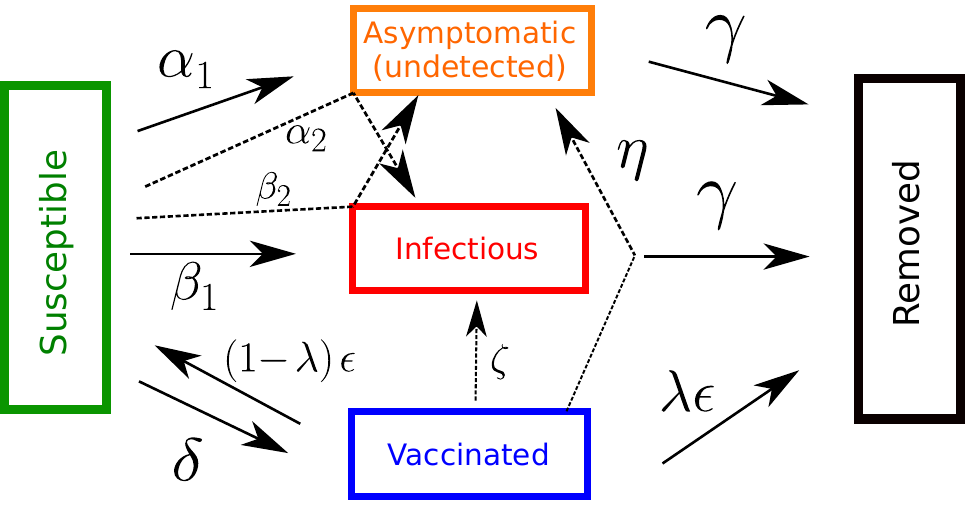}
\caption{
Illustration of the SAIVR model compartments and their inter-dependencies denoted by incoming and outgoing arrows and relevant flow parameters.
}
\label{fig:SAIVR_schematic}
\end{center}
\end{figure*}

The model involves two positive parameters, $\beta$ and $\gamma$ which govern the flow from one compartment to the other:\\
- $\beta$ is the transmission rate or effective contact rate of the disease: 
an infected individual comes into contact with $\beta$ other individuals per unit time 
(the fraction that are susceptible to contracting the disease is $S/N$); \\
- $\gamma$ is the removal rate. $\gamma^{-1}$  is the mean 
number of days who is infected spends in the Infectious compartment.\\
The SIR model obeys the following system of ordinary differential equations (ODE):
\begin{subequations}
\label{eq:SIR}
\begin{align}
\frac{dI}{dt} &= \beta I \frac{S}{N} - \gamma I  
\label{eq:SIR_I}
\\
\frac{dS}{dt} &= - \beta I \frac{S}{N} 
\label{eq:SIR_S}
\\
\frac{dR}{dt} &= \gamma I 
\label{eq:SIR_R}
\end{align}
\end{subequations}

Although the SIR model has been adopted
to study epidemic outbreaks in many previous works \cite{Saito_SIR, Fang_SIR, Smirnova_SIR, Alanazi_SIR, Palladino_SIR, Cooper_SIR, Kaxiras_2020}, it lacks few important aspects of the current ongoing pandemic.
First of all, it has been reported ~\cite{Asymptomatic_1, Asymptomatic_2} that an important fraction of those who are carrying the virus is asymptomatic. Since they often avoid contact tracing due to the absence of symptoms, they can spread the disease while remaining undetected.
 Furthermore, in December 2020 a global vaccination campaign has started. Vaccinating is a safe way to transfer people from the Susceptible to the Removed compartment bypassing the Infectious one thus reducing the likelihood of an outbreak.\\

The SAIVR model extends the SIR model by incorporating the two aforementioned   additional compartments:
\begin{itemize}
\item
$A(t)$ is the Asymptomatic/Undetected compartment that counts the number of those individuals that despite being infected are not tested/traced. This mainly occurs to those who recover from the infection without suffering any symptoms.

\item
$V(t)$ is the Vaccinated compartment. It takes into account those that have received a vaccine shot but are still not fully immunized by it.

\end{itemize}

The SAIVR model ODEs read:
\begin{subequations}
\label{eq:SAIVR}
\begin{align}
\frac{dI}{dt} &= \beta_1 I \frac{S}{N}  + \alpha_2 A \frac{S}{N} + \zeta I \frac{V}{N} - \gamma I
\label{eq:SAIVR_I}
\\
\frac{dA}{dt} &=  \alpha_1 A \frac{S}{N}   + \beta_2 I \frac{S}{N} + \eta A \frac{V}{N} - \gamma A
\label{eq:SAIVR_A}
\\
\frac{dS}{dt} &=   - \beta I \frac{S}{N} - \alpha A \frac{S}{N} - \delta \frac{S}{N} + (1-\lambda) \epsilon V
\label{eq:SAIVR_S}
\\
\frac{dV}{dt} &=  \delta \frac{S}{N}- \eta A \frac{V}{N} - \zeta I \frac{V}{N}  - \epsilon V
\label{eq:SAIVR_S}
\\
\frac{dR}{dt} &= \gamma I + \gamma A + \lambda \epsilon V
\label{eq:SAIVR_R}
\end{align}
\end{subequations}

The compartment inter-dependencies and flow are presented in Fig. \ref{fig:SAIVR_schematic}. 
The parameters of the SAIVR model  are the following:\\
- $\beta_1$ describes the rate at which individuals are exposed to symptomatic infection. An infected symptomatic individual comes into contact and infects $\beta_1$ susceptible individuals per unit time; \\
- $\alpha_1$ is the asymptomatic infection rate. An infected asymptomatic individual comes into contact with $\alpha_1$ susceptible individuals per unit time;  \\
- $\beta_2$ describes the rate at which susceptible individuals become asymptomatic infected after entering in contact with a symptomatic individual; \\
- $\alpha_2$ describes the rate at which who's susceptible becomes symptomatic after entering in contact with an asymptomatic individual; \\
- $\gamma$ retains the same meaning as in the SIR model, representing the mean removal rate.  $\gamma^{-1}$ is the mean 
amount of time individuals spend either in the Infectious or Asymptomatic compartments;\\
- $\zeta$ is the rate at which a vaccinated (but still not immune) individual enters in contact with a symptomatic infectious; \\
- $\eta$ describes the transmission rate at which who's asymptomatic comes into contact and infects vaccinated (but still not immune) individuals; \\
- $\delta$ is the first shot vaccination rate;\\
- $\lambda$ is the vaccine efficacy; \\ 
- $\epsilon^{-1}$ is the mean 
amount of time an individual spends in the Vaccinated compartment before reaching immunity and moving to the Removed compartment.\\\\
Countries and states do not respond to the disease as static entities passively facing the pandemic. They react by actively imposing (and relaxing) restrictive measures, learning how to effectively treat the infected, adjusting social interactions and by launching vaccination campaigns. Finally, the virus itself evolves in more infectious variants ~\cite{2021_variants}.\\
Country-specific parameters can be obtained by fitting the SAIVR model to a selected infectious wave occurred in a given country.
SAIVR has 14 adjustable parameters or initial conditions that needs to be estimated; given the scarcity of data (only the infectious and vaccinated populations are known) optimizing them presents a challenging problem. To address this, we either fixed some of them or employed a novel fitting method based on semi-supervised neural networks, which we present in the following section.

\section{Solving the SAIVR model with Neural Networks}\label{sec:NN}

\begin{figure*}[h]
\begin{center}
\includegraphics[width=1.0\textwidth]{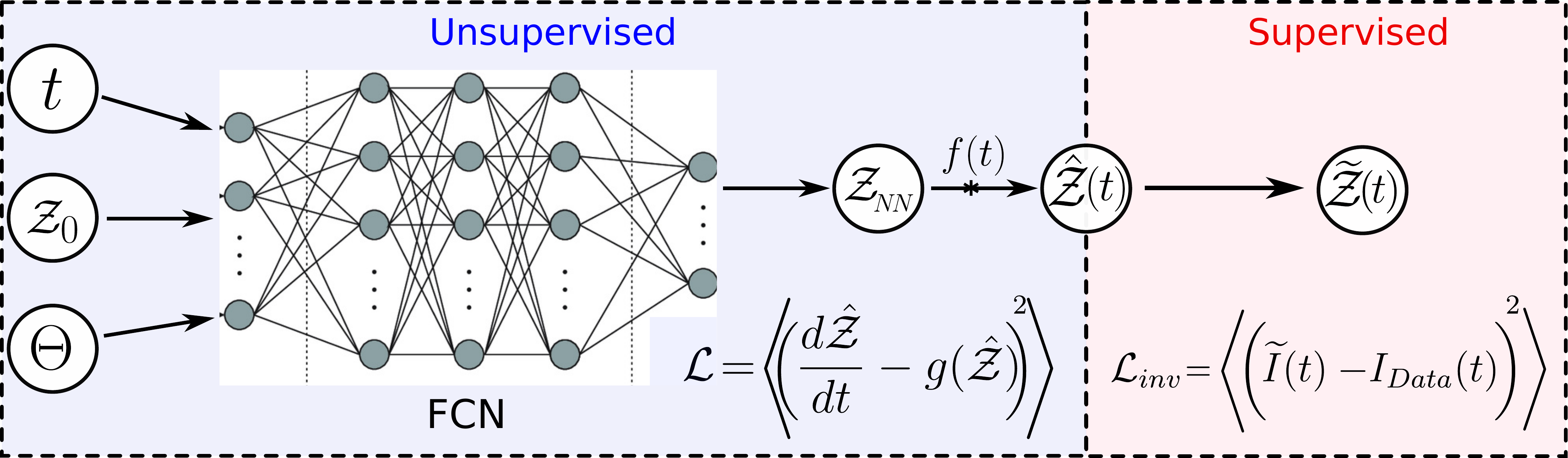}
\caption{ Semi-supervised network architecture. During the unsupervised procedure (blue box), a time sequence $t$, a set of initial conditions $\mathcal{Z}_0$ and parameter bundles $\Theta$ are fed as an input to a 6 layers fully connected network (FCN).
The output of the network $\mathcal{Z}_{NN}$ is multiplied by a function $f(t)$ to become a tentative parametric solution $\hat{\mathcal{Z}}(t)$ of the system of ODE in Eq.~\ref{eq:SAIVR}. 
The quality of $\hat{\mathcal{Z}}(t)$ is probed by the loss function $\mathcal{L}$.
When the network has learned the solutions, the inverse problem is then solved (red box).
An optimization algorithm selects the initial conditions and parameters in the bundle that best fit a given data-set. The loss $\mathcal{L}_{inv}$ depends on the infectious population of a given country/state $I_{Data}$ during the time sequence $t$.
}
\label{fig:NN_sketch}
\end{center}
\end{figure*}

In order to apply the SAIVR model we need a realistic estimate of the parameters and initial conditions for the system of Eqs.~\ref{eq:SAIVR}.
To obtain them we employed machine learning, a powerful method which has been extensively used for disease modeling \cite{Amazon, Yang_SEIR, Zou_NN, Paticchio_NIPS} and dynamical system forecasting \cite{Gallinari_NN, Chen_NN,Flamant}.
Our approach employs a semi-supervised procedure which determines the optimal set of initial conditions and parameters of the SAIVR model, yielding solutions that best fit a given data-set.
A sketch of this procedure is shown in Fig.~\ref{fig:NN_sketch}.

\subsection{Unsupervised learning} \label{sec:unsupervised}
The unsupervised part (blue box) consists of a data-free Neural Network (NN) that is trained to discover solutions for an ODE system of the form:
\begin{equation}
    \frac{d\mathcal{Z}}{dt} = g(\mathcal{Z}), \;\;\;\;\;\;\;\;\;\;\;\;\;\;\;\;\; \mathcal{Z}(t=0) = \mathcal{Z}_0
\end{equation}
where $\mathcal{Z} = (S(t),A(t),I(t),V(t),R(t))$ and $g(\mathcal{Z})$ is given in Eqs.\ref{eq:SAIVR}.
The NN takes as an input a time sequence $t$, a set of initial conditions $\mathcal{Z}_0$, and modeling parameters $\Theta$.\\
As we'll see in the following, $t$ is the set of days involved in a given epidemic wave going from $t_0$ to $t_0 + \Delta t$, $Z_0$ are the initial compartment populations and $\Theta$ some parameters of the SAIVR model.
The initial conditions and parameters are randomly sampled at each iteration $n$ over predefined intervals called bundles \cite{Flamant, Paticchio_NIPS}, so that the network learns an entire family of solutions.
The inputs propagate through the network until an output vector $\mathcal{Z}_{NN}$ of the same dimensions as the target solutions $\mathcal{Z}$ is produced.
The learned solutions $\hat{\mathcal{Z}}$ satisfy
the initial conditions identically by considering parametric solutions of the form:
\begin{equation}
\hat{\mathcal{Z}} = \mathcal{Z}_0 + f(t) (\mathcal{Z}_{NN} - \mathcal{Z}_0   )
\end{equation}
where $f(t) =1- e^{-t}$ \cite{Marios_2020}.
The loss function:
\begin{equation}
\label{eq:loss_de}
\mathcal{L} = \bigg \langle \left( \frac{d\mathcal{Z}}{dt} - g(\mathcal{Z})\right) ^2 \bigg  \rangle
\end{equation}
solely depends on the network predictions averaged ($\langle .. \rangle$) over all the iterations $n$, providing an unsupervised learning framework. 
Time derivatives are computed using the automatic-differentiation and back-propagation techniques \cite{automatic_differentiation}.

\subsection{Fitting a dataset}
\label{sec:fitting_method}
Once the NN is trained to provide solutions for the system of Eq. \ref{eq:SAIVR}, its weights and biases are fixed and the trained network  is used to develop a supervised pipeline for the estimation of the initial conditions and parameters, leading to solutions $\tilde{\mathcal{Z}}(t)$ that fit given data.
This procedure is illustrated in the red box in Fig.~\ref{fig:NN_sketch}. 
A solution $\hat{\mathcal{Z}}$ is generated by the network starting from $\bar{\mathcal{Z}_0}$ and $\bar{\Theta}$ randomly selected in the bundles. A stochastic gradient descent optimizer then adjusts $\bar{\mathcal{Z}_0}$ and $\bar{\Theta}$ in order to minimize the loss function:
\begin{equation}
\label{eq:loss_inv}
\mathcal{L}_{inv} = \bigg \langle \left( \tilde{I}(t) - I_{Data}(t) \right) ^2 \bigg \rangle 
\end{equation}
where $I_{Data}(t)$ is the infectious population of a given country/state and $\tilde{I}(t)$ is its NN fit.

\subsection{Applying the method to real data}
\label{sec:fitting}
We used this method to reproduce the most recent Covid-19 waves in 27 countries or states.
To test the generality of the model we selected epidemic waves that occurred in a broad range of geopolitical conditions, restrictive measures, time periods and vaccination efforts.
The bundles and fixed parameters used during the unsupervised training of the network are listed in Table.~\ref{tab:parameters}.

We found that the model is weakly sensitive on the choice of most parameters and thus, we kept some of them fixed during the training. 
The value of the vaccine efficacy $\lambda$ is based on Refs.\cite{Polack, Baden_2020}, where a vaccine efficacy of $94.8\%$ and $94.1\%$ is reported for the Pfizer-BioNTech and Moderna mRNA vaccines.
The $V \rightarrow I$ and $V \rightarrow A$ rates $\zeta$ and $\eta$
are derived by considering the order-of-magnitude ratio of the infected individuals in the vaccinated and placebo cohorts of Ref.~\cite{Polack}, with
$\beta_2$ and $\alpha_2$ set to $0.001$ and $0.01$ respectively, which are in order-of-magnitude agreement with the aforementioned results of the clinical trials.
The $V \rightarrow R$ rate $\epsilon$ is the inverse time an individual takes to acquire vaccine protection after the first shot. 
We set it to $\epsilon^{-1} = 21$ to reflect the fact that the second shot of vaccines is usually administered about three weeks after the first one.

\begin{figure*}[h]
\begin{center}
\includegraphics[width=1.0\textwidth]{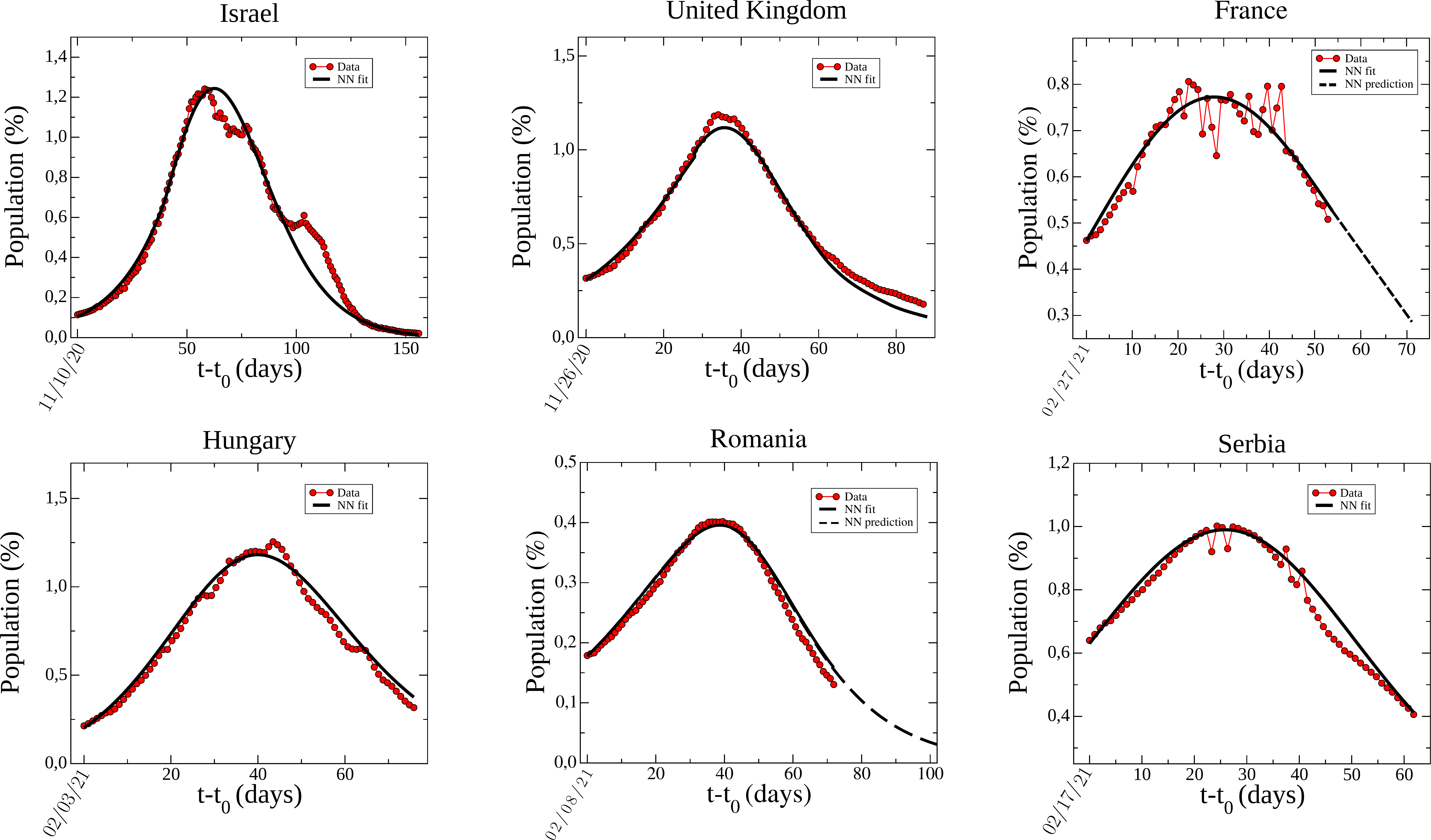}
\caption{ Infectious population percentage (red dots) of some selected countries in which a high percentage of their population has received a vaccine shot. The infectious population is expressed as a function of time (days). The date at which the wave began is pointed out on the horizontal axis. The neural network fits and predictions are shown by black solid and dashed lines respectively. 
}
\label{fig:vaccinated_countries_fit}
\end{center}
\end{figure*}

\begin{table*}
\begin{center}
\caption{Bundles and fixed parameters}
\begin{tabular}{cccccccc}
$I_0$ & $A_0$ & $V_0$ & $R_0$ & $\beta_1$ & $\gamma$ & $\alpha_1$ & $\delta$ \\
\midrule
$[0.1~\%, 2~\%]$ & $[0.1~\%, 2~\%]$ & $[0~\%, 60~\%]$ & $[0~\%, 30~\%]$ & $[0.1, 0.25]$ & $[0.07, 0.12]$  &  $[0.1, 0.25]$ & $[0, 0.03]$ \\
\end{tabular}

\begin{tabular}{cccccccc}
$\epsilon$ & $\lambda$ & $\eta$ & $\zeta$ & $\beta_2$  & $\alpha_2$  \\
\midrule
1/21 & 0.95 & 1e-2 & 5e-3 & 1e-3  &  1e-2  \\
\end{tabular}
\label{tab:parameters}
\end{center}

\end{table*}

\begin{figure*}[h]
\begin{center}
\includegraphics[width=1.0\textwidth]{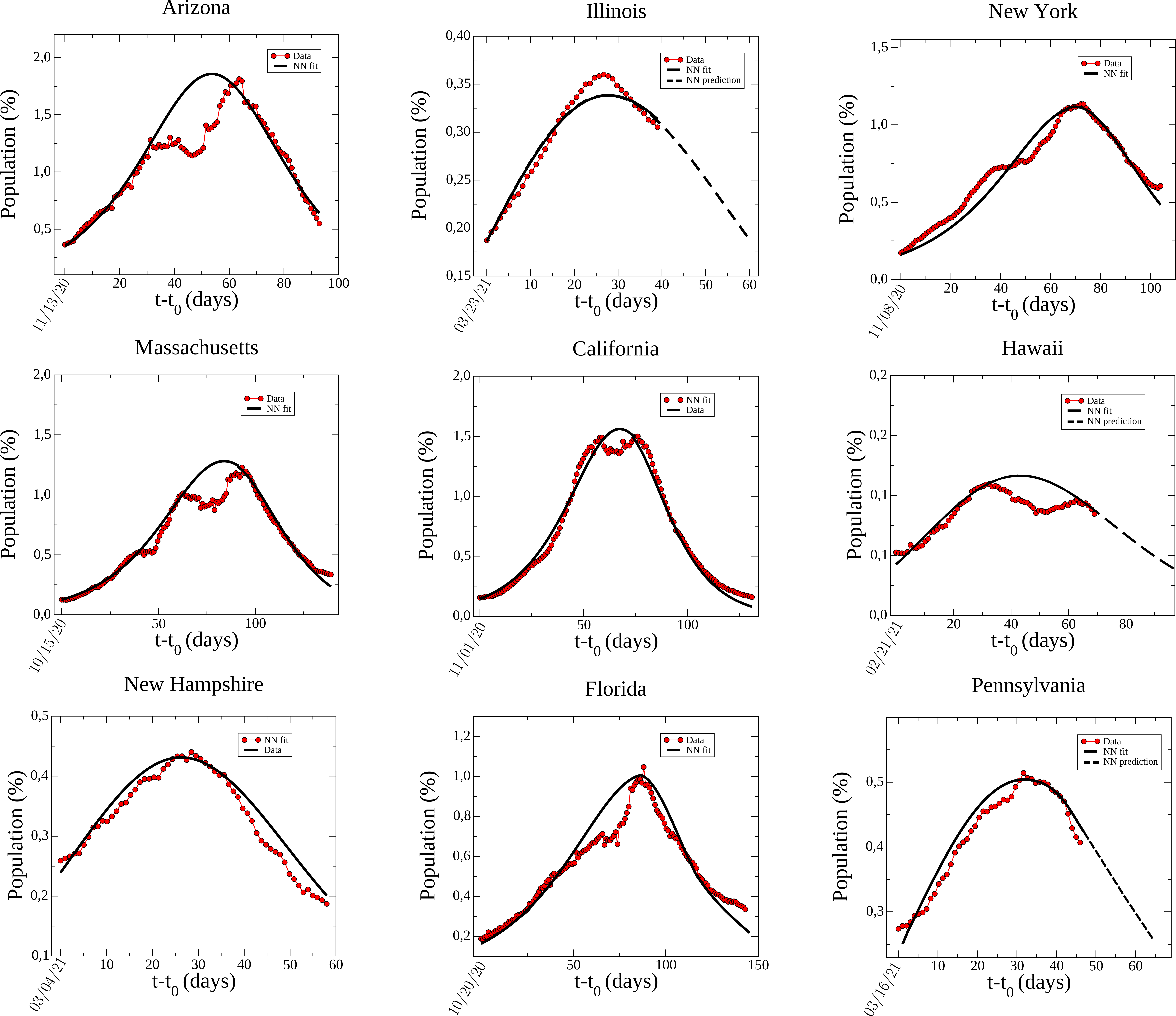}
\caption{
Infectious population percentage (red dots) of ten selected US states. The infectious population is expressed as a function of time (days). The date at which the wave began is pointed out on the horizontal axis. The neural network fits and predictions are shown by black solid and dashed lines respectively.
 }
\label{fig:US_states_fit}
\end{center}
\end{figure*}

The remaining parameters $\Theta = (\alpha_1, \beta_1, \delta, \gamma)$ strongly depend on what kind of restrictive measures are taken or how fast the vaccination campaign is, i.e. they are country dependent. We therefore selected them in bundles so that the network could learn solutions corresponding to a broad range of parameters and fit multiple countries.
The decay rate $\gamma$ is the inverse of the removal time which is about 1-2 weeks \cite{Recovery_time}.
The main symptomatic infection rate $\beta_1$ was sampled in an interval consistent with previous reports \cite{Kaxiras_2020}.
The earlier estimates that $80\%$ of infected population is asymptomatic has been considered
too high and have since been revised down  ~\cite{Asymptomatic_1, Asymptomatic_2}; the initial studies estimating this proportion were
limited by heterogeneity in case definitions, incomplete symptom assessment,
and inadequate retrospective and prospective follow-up of symptoms. We selected the main asymptomatic infection rate $\alpha_1$ to be varying in the same interval of $\beta_1$.  The first shot vaccination rate $\delta$ was selected based on known vaccination reports (see \ref{sec:matsmethod}).\\
Finally, the initial condition bundles $\mathcal{Z}_0 = (S_0,A_0,I_0,V_0,R_0)$ are defined over broad intervals able to cover the expected ($S,A,I,V,R$) populations at any given time for all the cases considered. Although the initial infected population $I_0$ is known, we still included it in the set of quantities to be fit by the network. We found that by doing so, the network generalizes better improving the fit of a given epidemic wave.

We then performed the fitting procedure described in \ref{sec:fitting_method} using the infectious populations of 27 countries/states. The data (including the number of vaccine shots administered) is retrieved from the `Our World in Data' GitHub repository \cite{OWID}.
Strictly speaking, the Infectious population of the SAIVR model is the amount of people that are actively infected by the virus on a given day, a number that should not be confused with the daily new cases. As such, it was computed as the difference between the total number of cases and of recovered/dead individuals.

We first applied the method to study the most recent Covid-19 wave in  some of the countries with the fastest vaccination campaigns and that managed to inoculate the first shot in at least $30\%$ of their population.
Fig.~\ref{fig:vaccinated_countries_fit} presents real data (red points), fits (black solid line), and some predictions (black dashed line) for the infectious populations of Israel, UK, Hungary, France, Romania and Serbia.
We also studied the USA, although due to the large size of the country we focused only on the largest states or those with the highest vaccination rates in the first quarter 2021, see Fig.~\ref{fig:US_states_fit}.
To assess the generality of the model and the fitting procedure, we applied it to other 12 countries spread throughout the world and which had at the end of spring 2021 a high number of cases.
The corresponding fits are shown in Figs. S1-S2 of the \textbf{Supplementary Material}.
As can be seen, the model is able to well reproduce all these epidemic curves despite missing some abrupt and rapid events that can be captured by more sophisticated multiple-wave models \cite{Kaxiras_2020}.
All the parameters determined by the network can be found in the \textbf{Supplementary Material};   their values are  within the bundles of Table.~\ref{tab:parameters}.
In particular, $\gamma$ was found in the $~ 10-12$ days range, $\beta_1$ oscillating in  $[0.14-0.19]$ while alpha was more volatile.
\\

\section{Insights on the future: vaccine hesitancy, herd immunity and new variants}

\begin{figure}[h]
\includegraphics[width=1.0\textwidth]{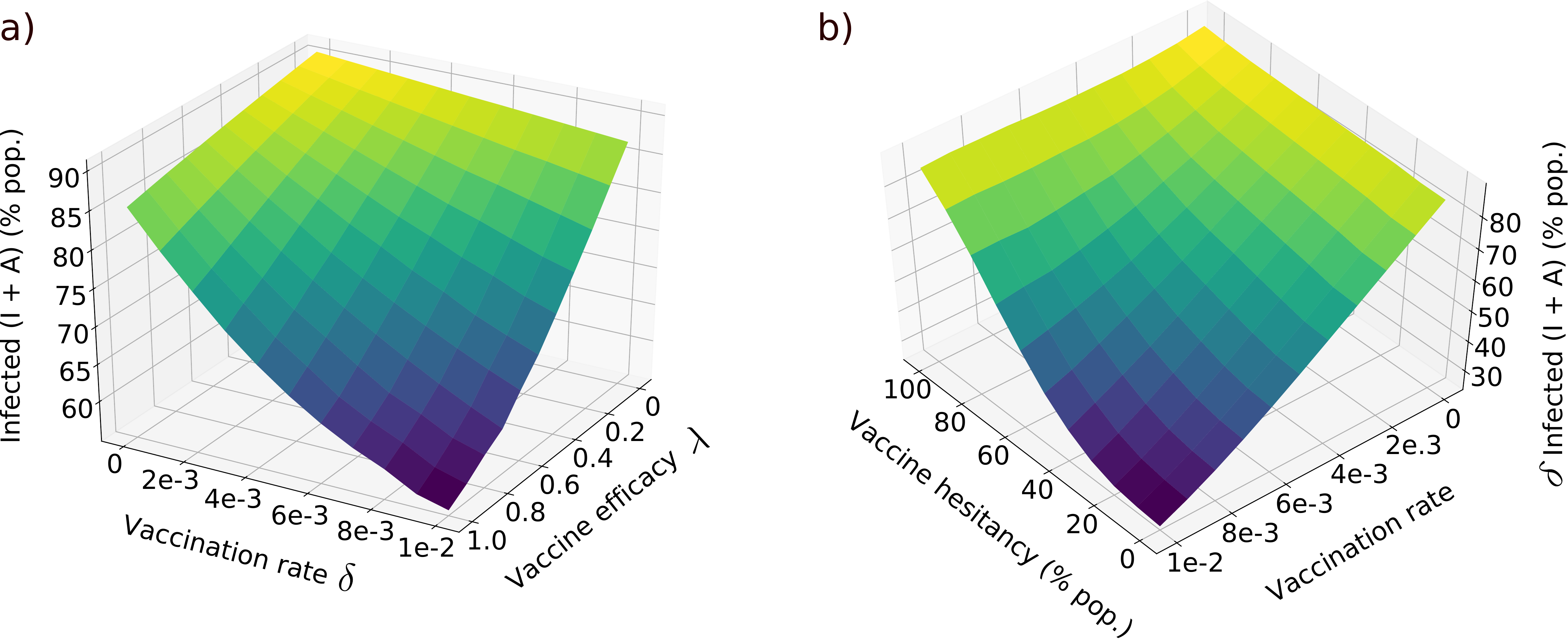}
\caption{Total infected population as a function of vaccination rate,  vaccine efficacy and vaccine denial population percentage. Results are obtained by numerically solving the SAIVR model for $I_0 = 10^{-5} N_{pop}$ and $A_0 = 0.2 \times I_0$, where $N_{pop} = 10^6$.
The parameters of the model used are those obtained by applying machine learning on the epidemic curves. \\
a) Infected population vs. vaccination rate $\delta$ and vaccine efficacy $\lambda$.\\
b) Infected population as a function of the percentage of the population that avoids getting vaccinated and the vaccine rate $\delta$.
}
\label{fig:hesitancy}
\end{figure}

\label{sec:future_scenarios}
In this section we use the results of the analysis performed in the previous section to study how the vaccination campaign is affecting the pandemic and its future evolution.
Unless otherwise specified, we set $\beta_1 = 0.16$, $\alpha_1 = 0.2$ and $\gamma = 1/12$, the average values retrieved from fitting real data. 
We start by pointing out how the vaccine efficacy is a key factor in halting the spread of the virus and how hesitancy is challenging the vaccination campaigns.
Finally, we discuss the concept of herd immunity and how it is affected by more infectious Covid-19 variants.

\subsection{Vaccination efficacy and hesitancy}

Figure \ref{fig:hesitancy} presents the total infected ($I + A$) population under increasing values of vaccination onset times ($T_0$), vaccination daily rates ($\delta$), vaccine efficacy ($\lambda$) and of vaccine hesitancy/denial population percentage. 
In the top panel, total infected population  is shown as a function of the vaccination rate $\delta$ and vaccine efficacy $\lambda$. As can be seen, even vaccines with a relatively low efficacy can rapidly reduce the infected population.
In Fig.~\ref{fig:hesitancy} b) we show how the number of those infected evolves as a function of $\delta$ and the percentage of population that avoids getting vaccinated.
These findings suggest that vaccine hesitancy, which accounts for a significant proportion of the population might seriously threaten the reach of herd immunity, especially if the situation is worsened by the appearance of more infectious Covid-19 strains.

\subsection{Herd immunity and new Covid-19 variants}

\begin{figure*}[h]
\includegraphics[width=1.0\textwidth]{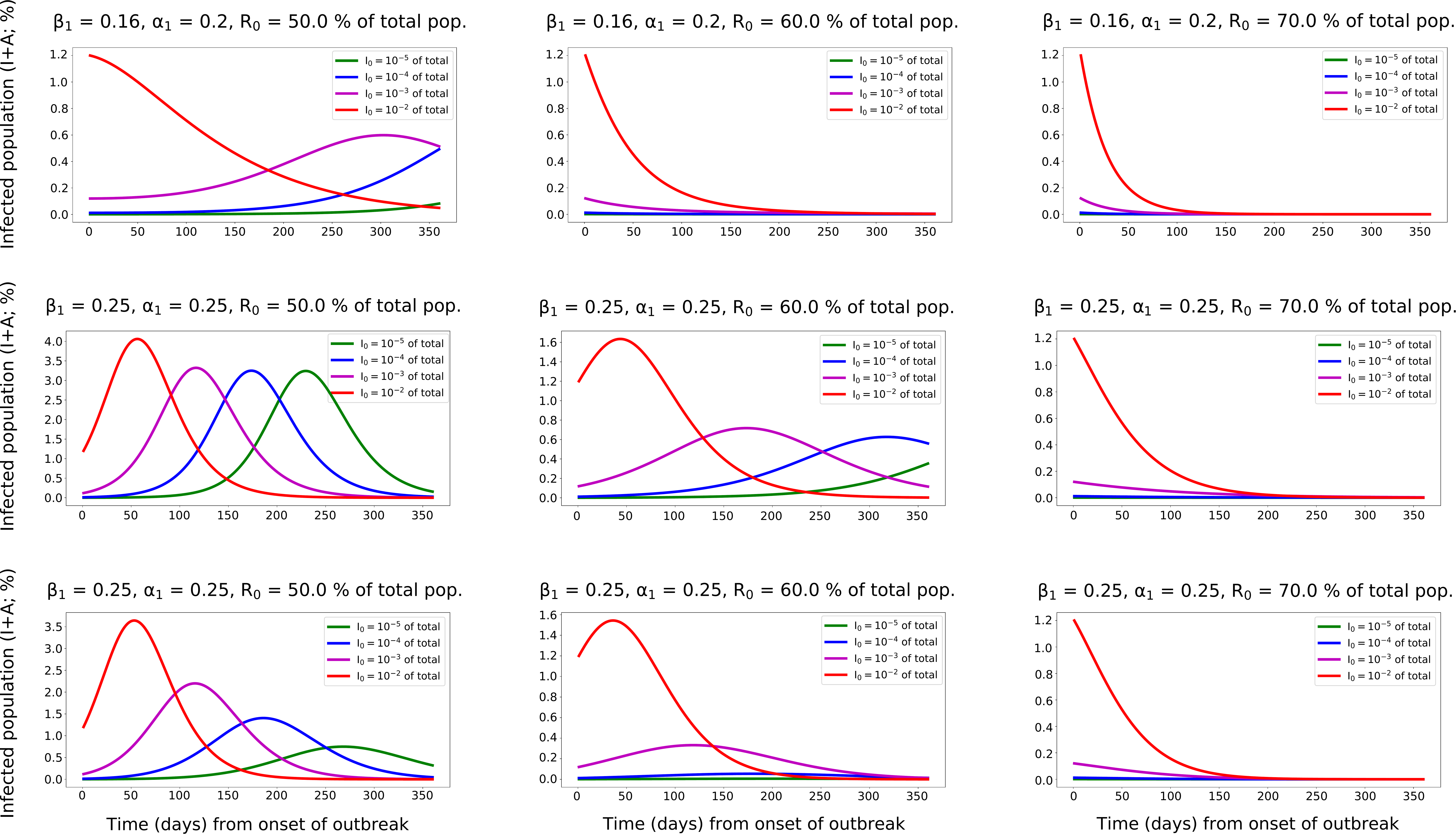}
\caption{
{\bf Top row}: Time evolution of epidemics following the introduction of infected individuals in a population that has been already vaccinated at 50$\%$ (left panel), 60$\%$ (middle panel), and 70$\%$ (right panel), with permanent immunity and no further vaccine roll-out, for four different numbers of newly infected individuals ($I_0 = 1 - 0.001 \%$ of total population, the color code is presented in the legends). Results are obtained from numerically solving the SAIVR model for $\beta_1 = 0.16$, $\alpha_1=0.20$, which represent the average of the countries' fitted values (see text for details and for the values of the other parameters). 
{\bf Middle row}: Same as in top row but with $\beta_1 = 0.25$, $\alpha_1=0.25$, which represent a more contagious Covid-19 variant. As is shown, vaccinated coverage of $50\%$ and $60\%$ cannot prevent the resurgence of outbreaks.
{\bf Bottom row}: Same as in middle row, but with ongoing vaccination roll-out with rate $\delta = 0.001$, per day, as it is shown, a continuing vaccine roll-out lowers the intensity of the resurgent waves and prevents the resurgence of subsequent outbreaks.
}
\label{fig:future_scenarios}
\end{figure*}

The achievement of herd immunity has been hailed as the ultimate goal of a successful vaccination campaign.
`Herd immunity', also known as `population immunity', is the indirect protection from an infectious disease that happens when a sufficient portion of the population is immune either through vaccination or immunity developed through previous infection.
Once the herd immunity threshold is met, the spread of the infectious disease is kept under control, current outbreaks will extinguish and endemic transmission of the pathogen will be interrupted. 
Earlier estimates of the threshold found values of about $60-70 \%$ of the population \cite{Howard_herd,Chowdhury2020,Aguas2020}. 
In reality, highly transmissible strains tend to increase the threshold value, possibly keeping this goal out of reach.
Furthermore, persistent hesitancy about vaccines makes vaccinating more than the $60 - 65\%$ of the population unlikely even in countries which are at the global forefront of the vaccination effort.

We quantitatively investigate the likelihood of incurring resurgent Covid-19 epidemics after having immunized $50\%$, $60\%$, and $70\%$ of the population, under different new infection introductions, Covid-19 variants and ongoing vaccine deployment pace.
Herd immunity protection is affected by the initial value of the removed population ($R_0$ at $t=0$), which comprises both recovered as well as fully vaccinated individuals, assuming permanent immunity for both cases. In each scenario, we study  the epidemic evolution after the introduction of a cluster of newly infected individuals in the population. $I_0$ represents the newly infected load at $t=0$ ($I_0 = 1\%, 0.1\%, 0.01\%, 0.001\%$ of the total population).

In the first part of the study we considered the less infectious variants spreading during the 2021 spring by using the parameters retrieved in Sec.~\ref{sec:fitting}.
The second scenario involves a more infectious Covid-19 strain such as the Delta variant, which has been reported to be able to spread the virus more efficiently  \cite{Delta_variant}.
Finally, we explore cases that involve or not further (continuing) vaccine roll-outs. 

Figure ~\ref{fig:future_scenarios} presents the results obtained by numerically solving the SAIVR model for the aforementioned cases. The top row presents the time evolution of an outbreak in a population where the 50$\%$ (left panel), 60$\%$ (middle panel), and 70$\%$ (right panel) of the individuals have been immunized, for different numbers of initially infected individuals $I_0$. Results are obtained by solving the SAIVR model for $\beta_1 = 0.16$, $\alpha_1=0.20$, and $\gamma = 1/12$; the average values obtained for the countries considered in Sec.~\ref{sec:fitting} and listed in \textbf{Supplementary Material}.
As it can be seen, when the immune portion of the population is only $50\%$, the outbreaks are contained but not eradicated as the virus spreads in low intensity waves 
making the disease endemic. Given the contagiousness of the less infectious variants, an immunity threshold larger than $60\%$ is enough to eradicate the disease.\\
The middle row of Fig.~\ref{fig:future_scenarios} presents the evolution of outbreaks driven by a more contagious variant; here $\beta_1 = 0.25$, and $\alpha_1=0.25$. As it is shown, if the immunized portion of the population is only $50\%$ or $60\%$, the resurgence of outbreaks cannot be prevented ($60\%$ immunity protection makes the disease endemic). Only when the $70\%$ of the population is immunized the disease is eradicated. \\
In both the top and the middle rows, the vaccine deployment is not taking place during the outbreaks. The bottom row instead considers the highly infectious variant disease evolution but with constant vaccine roll-out (with rate $\delta = 0.001$).
 As it can be seen, since even after getting only the first vaccine shot individuals are partially protected, continuing the vaccine administration rapidly lowers the intensity of the resurgent waves and helps preventing subsequent outbreaks. 

Although recent reports on highly infectious variants claim that the efficacy of most vaccines is still about $90 \%$ in preventing serious illnesses ~\cite{Delta_vaccine}, it is still not clear their performance on halting asymptomatic transmission.
In this study, we assumed that the vaccine efficacy on protecting from more infectious variants is the same as for the less infectious ones.
Despite this optimistic assumption, the herd immunity threshold is moved to higher values by simply increasing the infection rates.

\section{Discussion} \label{sec:Conclusions}
Compartmental models are efficient tools to deal with the time evolution of disease outbreaks. They provide us with useful intuition on the impact of non-pharmaceutical intervention in decreasing the number of infectious incidence rates.\\
In this work, we have augmented the classic SIR model with the ability to accommodate asymptomatic transmission and vaccinated individuals.
The SAIVR model is a straightforward deterministic model, which does not take into consideration age, gender or geographic clustering.
Despite this, its simplicity and the insights it offers on how key epidemiological variables affect individuals are among its main strengths. 
Its power also lies in the fact that, as factors such new variants are added to the model, it is easy to adjust its parameters and provide with best fit curves between the data and the model predictions.\\
Since the inclusion of the Asymptomatic and Vaccinated compartments enlarged the number of parameters and initial conditions of the model, we employed a novel semi-supervised framework to estimate most of them.
An unsupervised neural network solves the model's differential equations over a range of parameters and initial conditions. A supervised approach then incorporates data and determines the optimal initial conditions and modeling parameters that best fit the 27 epidemic curves considered.
As expected due to the heterogeneity of the countries sample,
the resulting parameters fit are dissimilar although they follow similar trends.\\
We used these results to shed light on the impact of the vaccination campaign on the future of the pandemic.
We pointed out how vaccine hesitancy is one of the most important hurdles of the campaign and further efforts should be done to support people and give them correct information about vaccines. 
Because of this, vaccinating the critical number of people that have to be immune in order to prevent future outbreaks (i.e. herd immunity), is likely to be out of reach. Widely circulating coronavirus variants are also a threat as they move the herd immunity threshold to higher values.
This points out the importance of rapidly reducing the infection rate by any means, such as
by imposing restrictive measures in case highly infective new variants appears before the herd immunity threshold is reached.
These results manifest the need for continuing the vaccination effort and the drive for achieving high vaccination coverage in order to contain outbreaks generated by new and possibly more infectious variants.\\

{\bf Data availability}\\
The code used to perform the fitting is available on GitHub \cite{Github_Mattia}. All study data are either included in the article and
supporting information or available in Ref.~\cite{OWID}. \\

{\bf Author Contribution}\\
E.K. conceived the proposed model and supervised the study.  M.A, G.N. and M.M. designed the code. M.A. and G.N. performed numerical experiments, collected data and analyzed the results. M.A. wrote the initial
draft of the manuscript. All authors critically revised, improved, and reviewed the manuscript in various
ways, and gave final approval for publication.\\

{\bf Competing interests}\\
The authors declare no competing interest.

\appendix

\section{Computational methods}
\label{sec:matsmethod}
We implemented a fully connected feed forward neural network that consists of six hidden layers with 48 neurons per layer and Sigmoid activation functions. The code is written in PyTorch \cite{automatic_differentiation} and published on GitHub \cite{Github_Mattia}. In the following we list the technical details of the learning procedure described in Sec.~\ref{sec:NN}:
\label{methods}
\subsection{Unsupervised learning}

The network parameters (weights and biases) are updated with an adaptive learning rate  using the Adam algorithm \cite{Adam} until the loss $\mathcal{L}$ of Eq. \ref{eq:loss_de} becomes smaller than $10^{-8}$. The learning rate ranged from 0.001 to $10^-6$.
The (initial) infectious  and asymptomatic populations involved in realistic situations are  a small fraction of the total population. 
The ODEs in Eqs. \ref{eq:SAIVR} are highly non-linear and extremely sensitive on initial conditions. Subsequently,  training the unsupervised network for the initial conditions given in Tab. \ref{tab:parameters} is very challenging.
To cope with this, we started by training the network with a different choice of initial condition bundles taken in `safer' intervals ($I_0, A_0 = [10\%, 30\%]$) where the network was able to quickly learn the solutions with high-accuracy.
Then, we gradually decreased $I_0$ and $A_0$ until they meet the values in Tab. \ref{tab:parameters}, while we keep training the network using the previously trained weights and biases.  
After performing this annealing procedure, the network was able to learn the solutions of the family of ODE defined by Tab.~\ref{tab:parameters}.

\subsection{Supervised learning and fitting procedure}
The SAIVR parameters  and initial conditions are updated with a stochastic gradient descent algorithm.
Whenever the infection wave started before the vaccination campaign, we divided the time series in two parts: before ($\delta = 0$) and after ($\delta > 0$) the vaccination campaign began. 
The first shot vaccination rate $\delta$ can abruptly change in time based on logistics, social or political reasons. 
Therefore, instead of fitting $\delta$ as we did with the other parameters in the bundles we estimated it using real data, with the aim of reproducing the average number of vaccinated people.
The Vaccinated compartment counts those individuals who have received the first shot but have not been fully immunized yet.
We defined the average number of people in the vaccinated compartment  $\tilde{V}$ in a given time interval as $\tilde{V} = \int_{t_0}^{t_f} V(t) \epsilon~ dt$, where $\epsilon$ is the $V \rightarrow R$ rate.
Then, we computed with a similar procedure $\tilde{V}_{Data}$ using data relative to a given country and where $V_{Data}(t)$ is the difference between the total number of people who got at least one shot and those fully vaccinated.
$\delta$ and $V_0$ were selected to match $\tilde{V}$ with $\tilde{V}_{Data}$.\\
Although $I_0$ was provided in the data, we did not fix it during the fitting. We have
empirically observed that the network generalizes better in this case.\\
The fitting procedure was performed 20 times for any given data-set in order to select the parameters and initial conditions that fit the ground-truth data with the lowest loss of Eq. \ref{eq:loss_inv}.






\end{document}